\documentclass[prl,
,tightenlines
,showkeys
,showpacs
,twocolumn
]
{revtex4-2}

\usepackage{amsmath,amssymb}
\usepackage{bm}
\usepackage[dvipdfmx]{graphicx}
\usepackage{ascmac} 
\usepackage{fancybox}
\usepackage{float}
\usepackage{enumerate}
\usepackage{color}
\usepackage{amsthm}
\usepackage{mathrsfs}
\usepackage{comment}
\usepackage{ulem}
\allowdisplaybreaks[1]
\usepackage{comment}
\newtheorem*{th.}{Theorem}

\newcommand{\vsigma}{\vec{\sigma}}

\begin{document}
	\title{Short composite rotation robust against two common systematic errors}
	\author{Shingo Kukita$^{1)}$}
	\email{toranojoh@phys.kindai.ac.jp}
	\author{Haruki Kiya$^{1)}$}
	\email{kiya.haruki@kindai.ac.jp }
	\author{Yasushi Kondo$^{1)}$}
	\email{ykondo@kindai.ac.jp}
	\affiliation{$^{1)}$Department of Physics, 
		Kindai University, Higashi-Osaka 577-8502, Japan}
	
	\begin{abstract}
		Systematic errors hinder precise quantum control.
		Pulse length errors (PLEs) and off-resonance errors (OREs) are typical systematic errors that are encountered during one-qubit control. 
		A composite pulse (CP) can help compensate for the effects of systematic errors during quantum operation.
		Several CPs that are robust against either PLE or ORE have been identified.
		However, few attempts have been made to construct CPs that are robust against both errors (bi-robust).
		We develop a novel bi-robust CP for one-qubit operations by modifying a PLE robust CP, 
		which exhibits a shorter operation time than that of previously developed bi-robust CPs. 
	\end{abstract}
	\maketitle
	
	Quantum information technologies, such as quantum computing \cite{bennett2000quantum,Nielsen2000,nakahara2008quantum}, quantum communication \cite{ekert1991quantum,gisin2007quantum,chen2021integrated}, and quantum metrology \cite{helstrom1976quantum,caves1981quantum,holevo2011probabilistic} have been receiving increasing attention.
	These technologies require the precise control of quantum states.
	However, systematic errors due to experimental apparatuses hinder precise control.
	Therefore, a method to compensate for the effects of systematic errors needs to be developed. One suitable means is the use of a composite pulse (CP) \cite{counsell1985analytical,levitt1986composite,claridge2016high}.
	CPs have been developed in the field of nuclear magnetic resonance (NMR)\cite{levitt2013spin}, where a simple model quantum computer can be realized \cite{gershenfeld1997bulk,cory1997ensemble,vandersypen2001experimental,JONES201191}.
	In CPs, a single operation is replaced by a sequence of operations such that the errors corresponding to different operations cancel each other.
	Hereinafter, we refer to this method as CP following the conventions in NMR; this method can also be called a composite quantum operation or composite quantum gate, which can be widely applied to quantum information.
	CPs have been validated in several fields of quantum information, such as NMR \cite{bando2020concatenated}, nitrogen-vacancy centers \cite{said2009robust}, superconducting qubits \cite{collin2004nmr}, and ion traps \cite{gulde2003implementation,timoney2008error}.
	
	In one-qubit control, two typical systematic errors are observed: pulse length error (PLE) and off-resonance error (ORE).
	PLE is caused by the deviation of the control field strength from the expected value, while ORE is caused by the miscalibration of the resonance frequency of the qubit being controlled.
	In particular, several PLE-robust CPs have been identified, such as SK1 \cite{brown2004arbitrarily}, BB1 \cite{wimperis1994broadband}, and SCROFULOUS \cite{cummins2003tackling}.
	A well-investigated ORE-robust CP is CORPSE proposed in Ref. \cite{cummins2000use}, which has been widely applied \cite{mottonen2006high,said2009robust,timoney2008error,bando2012concatenated}.
		
	However, limited studies have investigated CPs that are robust against both PLEs and OREs.
	(Hereinafter, we refer to this robustness against both PLEs and OREs as bi-robustness.)
	References \cite{jones2013designing} and \cite{ryan2010robust} discuss CPs performing a bi-robust operation of $\pi$-rotation in the Bloch sphere.
	However, these CPs only perform a $\pi$-rotation and cannot execute rotations at an arbitrary angle.
	Another technique, ConCatenated Composite Pulse (CCCP), is proposed in Ref. \cite{bando2012concatenated}.
	This method can construct a bi-robust CP by concatenating PLE- and ORE-robust CPs.
	We can realize bi-robust rotation at an arbitrary angle using CCCP.
	However, because CCCP concatenates two types of CPs to obtain bi-robustness,
	the total operation time tends to be long.
	The total operation time of a CP is also a key criterion that must be considered in the evaluation of its performance, because a longer operation time implies stronger decoherence caused by the environment surrounding the qubit.
	In addition, a shorter operation time is advantageous to precisely control the one-qubit state under interaction with other systems.
	
	We propose a bi-robust CP constructed based on a simple concept that is different from that of CCCP.
	We use SCROFULOUS \cite{cummins2003tackling}, a PLE-robust CP, as a ``seed'' to construct a bi-robust CP.
	We further decompose an operation in SCROFULOUS into several forward and backward operations to include ORE robustness to this sequence.
	We call this idea a {\it switchback} technique because the decomposed trajectory resembles a railway switchback.
	The resulting CP is bi-robust, and exhibits the shortest operation time among the existing bi-robust CPs.
	Therefore, the proposed bi-robust CP can be applied to any field relating to quantum information that requires bi-robustness and a short operation time.
	
	We consider the following family of unitary operations:
	\begin{equation}
		(\theta)_{\phi}=\exp \Bigl(-i \theta \vec{n}_{\phi}\cdot \frac{\vsigma}{2}\Bigr),~~\vec{n}_{\phi}=(\cos \phi, \sin \phi,0),
		\label{eq:unitary}
	\end{equation}
where $\vsigma=(\sigma_{x},\sigma_{y},\sigma_{z})$ denotes the vector comprising the Pauli matrices.
	In the Bloch sphere of one-qubit states, the above unitary operation represents a rotation with an angle $\theta$ and the axis $\vec{n}_{\phi}$ directed into the $xy$ plane.
	For simplicity, we assume that the CPs are constructed by sequentially performing elementary operations in the form (\ref{eq:unitary}).
	Our target operation is also assumed to be written in this form.
	These assumptions are common in NMR.
	When a one-qubit operation suffers from PLEs and OREs, the unitary (\ref{eq:unitary}) is deformed as follows:
	\begin{align}
		(\theta)^{(\epsilon,f)}_{\phi}:=&~\exp \Biggl(-i \theta (1+\epsilon) \Bigl(\vec{n}_{\phi}\cdot \frac{\vsigma}{2}+f\frac{\sigma_{z}}{2} \Bigr)\Biggr)\nonumber\\
		\sim&~(\theta)_{\phi}-i \epsilon \Bigl(\theta \vec{n}_{\phi}\cdot \frac{\vsigma}{2}\Bigr) (\theta)_{\phi}-i f \sin(\theta/2) \sigma_{z}\nonumber\\
		&~~~~~~~~~~~~~~~~~~~+{\cal O}(\epsilon^{2},f^{2},\epsilon f),
	\end{align}
where $\epsilon$ ($f$) is a small parameter representing the magnitude of the PLE (ORE).
	A (first-order) PLE-robust CP comprising $k$ elementary operations $(\theta_{k})_{\phi_{k}}\cdots(\theta_{1})_{\phi_{1}}$ satisfies
	\begin{equation}
		(\theta_{k})^{(\epsilon,0)}_{\phi_{k}}\cdots(\theta_{1})^{(\epsilon,0)}_{\phi_{1}}=(\theta)_{\phi}+{\cal O}(\epsilon^{2}),
		\label{eq:PLErobust}
	\end{equation}
	for the target operation $(\theta)_{\phi}$.
	Similarly, we define an ORE-robust CP.
	We aim to determine a bi-robust CP, which is defined as follows:
	\begin{equation}
			(\theta_{k})^{(\epsilon,f)}_{\phi_{k}}\cdots(\theta_{1})^{(\epsilon,f)}_{\phi_{1}}=(\theta)_{\phi}+{\cal O}(\epsilon^{2},f^{2},\epsilon f).
			\label{eq:bi-robust}
	\end{equation}
	To this end, we utilize the {\it switchback} technique explained later.
	
	First, we review the ``seed'' of a bi-robust CP, SCROFULOUS (short composite rotation for undoing length over and under shoot) \cite{cummins2003tackling}.
	SCROFULOUS is a $k=3$ symmetric PLE-robust CP determined as
	\begin{align}
		\theta_{1}=&~\theta_{3}={\rm arcsinc}\Bigl( \frac{2 \cos(\theta/2)}{\pi}\Bigr),\nonumber\\
		\theta_{2}=&~\pi,\nonumber\\
		\phi_{1}=&~\phi_{3}=\phi+\arccos\Bigl(-\frac{\pi \cos \theta_{1}}{2 \theta_{1} \sin(\theta/2)}\Bigr),
		\nonumber\\
		\phi_{2}=&~\phi_{1}-\arccos\Bigl(- \frac{\pi}{2\theta_{1}}\Bigr),
	\end{align}
	where the target operation is $(\theta)_{\phi}$ and ${\rm arcsinc}$ is the inverse of the ${\rm sinc}$ function.
	One can easily verify that SCROFULOUS is not ORE-robust but can be used as a ``seed'' to construct a bi-robust CP.
	We replace the second rotation $(\pi)_{\phi_{2}}$ with $(\theta_{r})_{\phi_{2}+\pi}(\pi+2\theta_{r})_{\phi_{2}}(\theta_{r})_{\phi_{2}+\pi}$, where $\theta_{r}$ is a parameter that can be varied to ensure the ORE robustness.
	This replacement maintains the PLE robustness of SCROFULOUS because each rotation and its error terms commute with each other and then
	\begin{equation}
		(\theta_{r})^{(\epsilon,0)}_{\phi_{2}+\pi}(\pi+2\theta_{r})^{(\epsilon,0)}_{\phi_{2}}(\theta_{r})^{(\epsilon,0)}_{\phi_{2}+\pi}=(\pi)^{(\epsilon,0)}_{\phi_{2}}.
	\end{equation}
	Thus, a composite sequence $(\theta_{r})_{\phi_{2}+\pi}(\pi+2\theta_{r})_{\phi_{2}}(\theta_{r})_{\phi_{2}+\pi}$ has an error-preserving property for PLE \cite{bando2012concatenated}.
	However, this replacement changes the ORE dependence of the rotation:
	\begin{equation}
		(\theta_{r})^{(0,f)}_{\phi_{2}+\pi}(\pi+2\theta_{r})^{(0,f)}_{\phi_{2}}(\theta_{r})^{(0,f)}_{\phi_{2}+\pi}\neq (\pi)^{(0,f)}_{\phi_{2}}.
	\end{equation}  
	Thus, we now obtain a free parameter $\theta_{r}$ to add the ORE robustness to the SCROFULOUS sequence while maintaining its PLE robustness.
	The proposed method is termed as a {\it switchback} technique.
	The reason for replacing $(\pi)_{\phi_{2}}$ with the symmetric sequence is explained in the Supplementary Material.
	
	We now obtain the $k=5$ symmetric PLE-robust sequence $(\theta_{1})_{\phi_{1}}(\theta_{r})_{\phi_{2}+\pi}(\pi+2\theta_{r})_{\phi_{2}}(\theta_{r})_{\phi_{2}+\pi}(\theta_{1})_{\phi_{1}}$.
	The ORE robustness condition of this sequence is obtained by considering the first-order term of $f$ to be zero, as in Eqs. (\ref{eq:PLErobust}) and (\ref{eq:bi-robust}).
	We determine that this condition is equivalent to the following equality:
	\begin{equation}
		\cos \theta_{r}=\frac{1}{2}\biggl(1- \frac{\pi (\sin(\theta_{1}/2))^{2}}{\theta_{1}} \biggr).
	\end{equation}
	(See Supplementary Material for the derivation). 
	When $\theta_{r}$ satisfies the above equality, the CP becomes ORE-robust.
	Thus, we obtain the $k=5$ bi-robust CP.
	We call this sequence Short COmposite Rotation BUffering Two Undesirables with Switchback (SCORBUTUS).
	
\begin{figure}[h]
		\begin{center}
			\includegraphics[width=88mm]{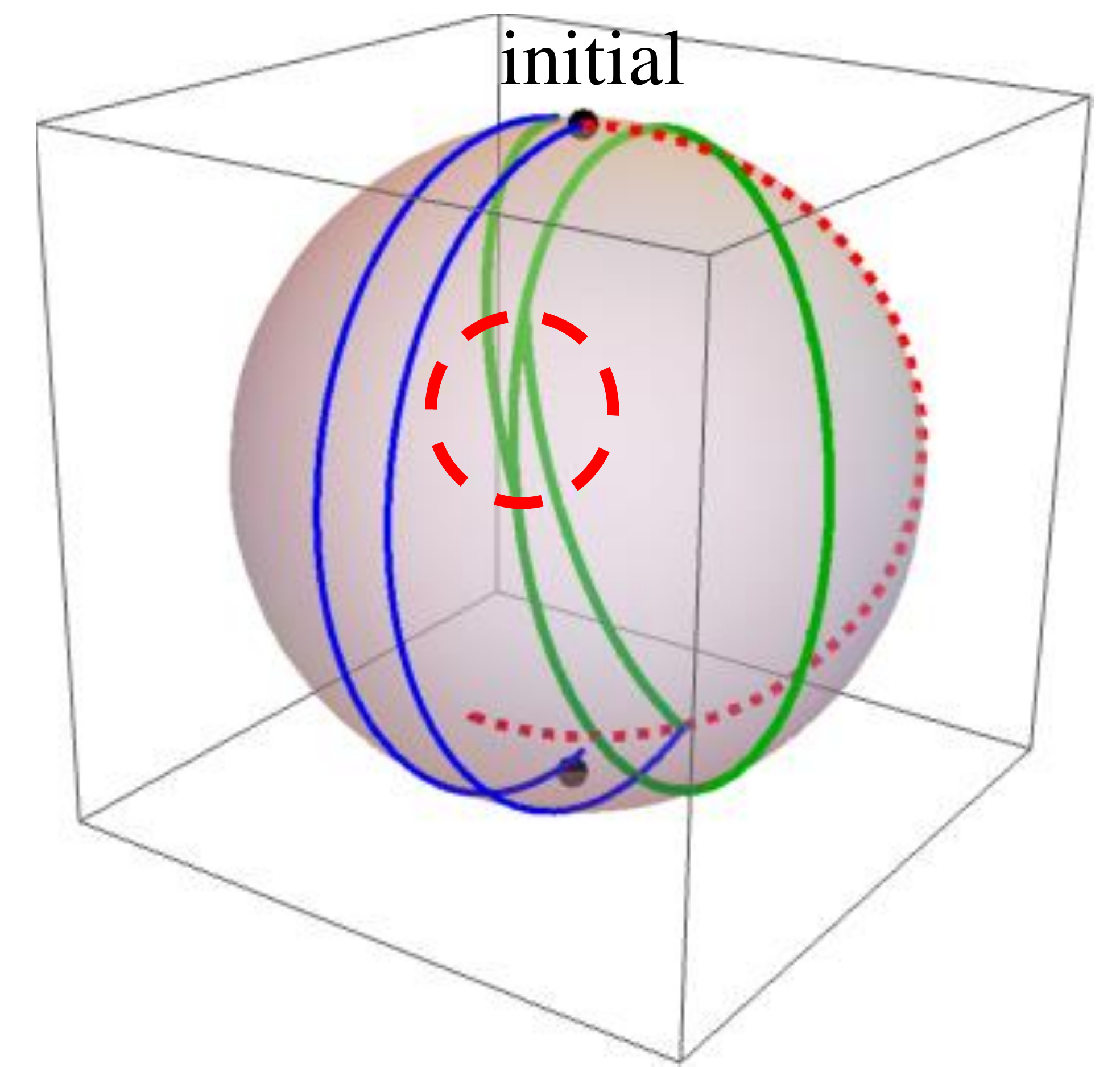}
			\caption{(color online) Trajectory of the elementary $(\theta)^{(\epsilon,f)}_{0}$ (dashed line) and the corresponding SCORBUTUS sequence (solid line) on the Bloch sphere.
			The green solid lines correspond to the second operation in the original SCROFULOUS sequence.
			The circle surrounds the switchback behavior during this operation.
			The initial state is considered to be the north pole, and the terminal point by the ideal operation is considered as the south pole. They are represented as dots.
			We assume the error parameters as $\epsilon=0.1$ and $f=0.1$.
			\label{fig:trajectory}
			}
		\end{center}
	\end{figure}

	In Fig.~\ref{fig:trajectory}, we depict the SCORBUTUS trajectory targeting $(\pi)_{0}$ with PLE and ORE on the Bloch sphere.
	The initial state represented by the north pole is transferred to a point close to the south pole by SCORBUTUS,
	while the elementary $(\pi)^{(\epsilon,f)}_{0}$ directly suffers from both PLE and ORE.
	We observe a zigzag motion (the green solid lines) resembling a railway switchback during the operation, which is why we refer to this as a switchback technique. 
	
	Here, we discuss a crucial advantage of SCORBUTUS, i.e., the total operation time.
	For the bi-robust CPs performing an arbitrary $(\theta)_{\phi}$ as the target operation, the one with the shortest operation time thus far is (reduced) SKinsC introduced in Ref. \cite{bando2012concatenated}.
	This is a $k=6$ bi-robust CP whose sequence is given as
	\begin{align}
		\theta_{1}=&~\theta_{5}=\theta_{6}=\frac{\theta}{2}-\arcsin\Bigl( \frac{\sin\theta/2}{2}\Bigr),\nonumber\\
		\theta_{2}=&~2\pi-\frac{\theta}{2}-\arcsin\Bigl( \frac{\sin\theta/2}{2}\Bigr),\nonumber\\
		\theta_{3}=&~\theta_{4}=2\pi,\nonumber\\
		\phi_{1}=&~\phi_{2}-\pi=\phi_{5}-\pi=\phi_{6}=\phi,\nonumber\\
		\phi_{3}=&~\phi+\pi-\arccos\Bigl(-\frac{2\pi-\theta}{4\pi} \Bigr),\nonumber\\
		\phi_{4}=&~\phi+\pi+\arccos\Bigl(-\frac{2\pi-\theta}{4\pi} \Bigr),
	\end{align}
	for the target rotation $(\theta)_{\phi}$.
	This sequence is slightly different from that presented in Ref. \cite{bando2012concatenated};
	SKinsC in Ref. \cite{bando2012concatenated} has several errata and the above one is correct.
	\begin{figure}[h]
		\begin{center}
			\includegraphics[width=88mm]{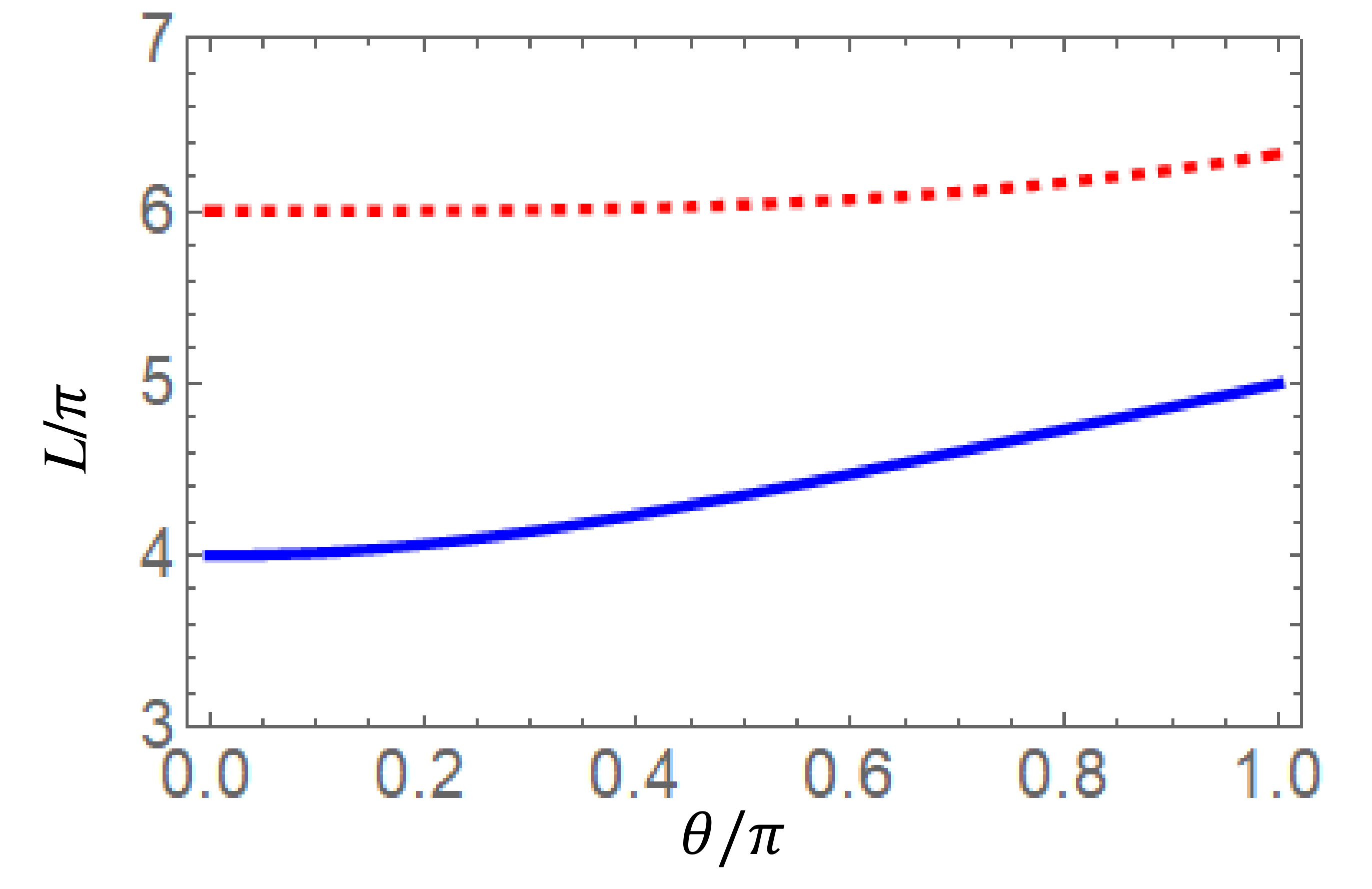}
			\caption{
				(color online) Total operation time $L$ as a function of $\theta$ for the target operation.
				The blue (red) line represents the total time of SCORBUTUS (SKinsC). 
				\label{fig:length}
			}
		\end{center}
	\end{figure}
	Fig.~\ref{fig:length} depicts the total (nondimensionalized) time of the operation, $L:=\sum^{k}_{i=1}\theta_{i}$ for $(\theta_{k})_{\phi_{k}}\cdots(\theta_{1})_{\phi_{1}}$, for SCORBUTUS and SKinsC as a function of $\theta$ in the target operation $(\theta)_{\phi}$.
	Note that $\phi$ does not change the total operation time.
	Evidently, SCORBUTUS has a shorter operation time than that of SKinsC for any target rotation angle $\theta$.
	Thus, to the best of our knowledge, the developed bi-robust CP has the shortest operation time.
	
\begin{figure*}[t]
		\begin{center}
			\includegraphics[width=175mm]{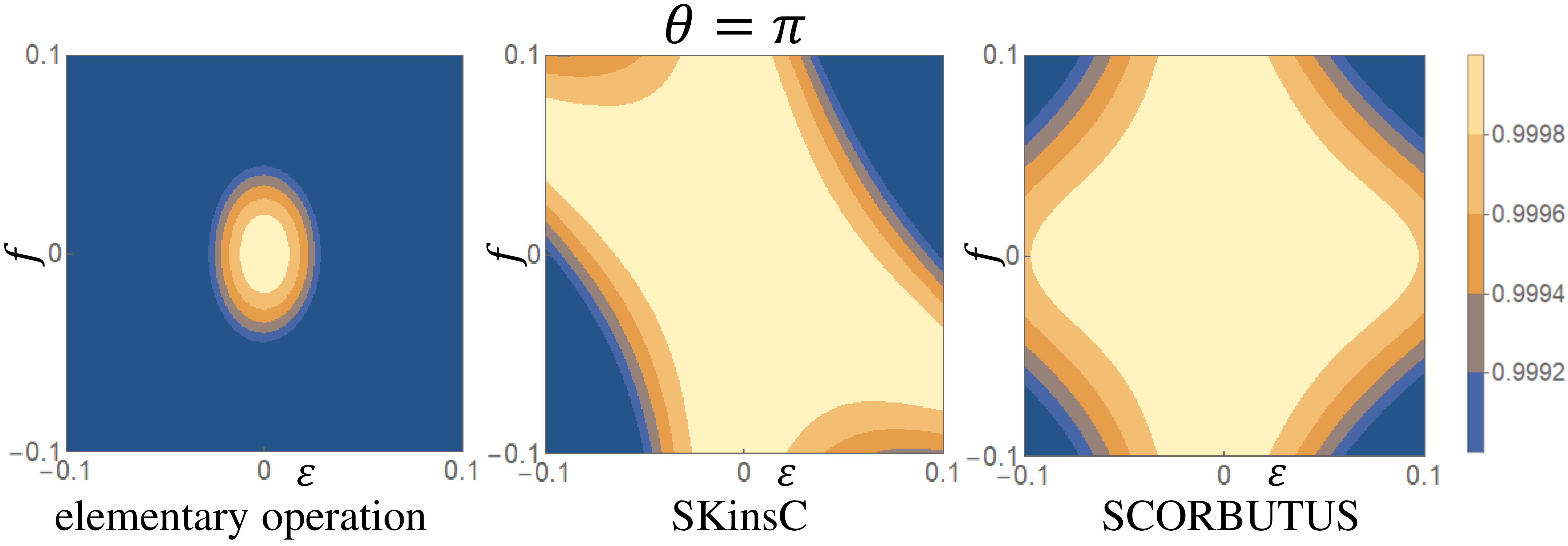}\\
			\vspace{0.3cm}
			\includegraphics[width=175mm]{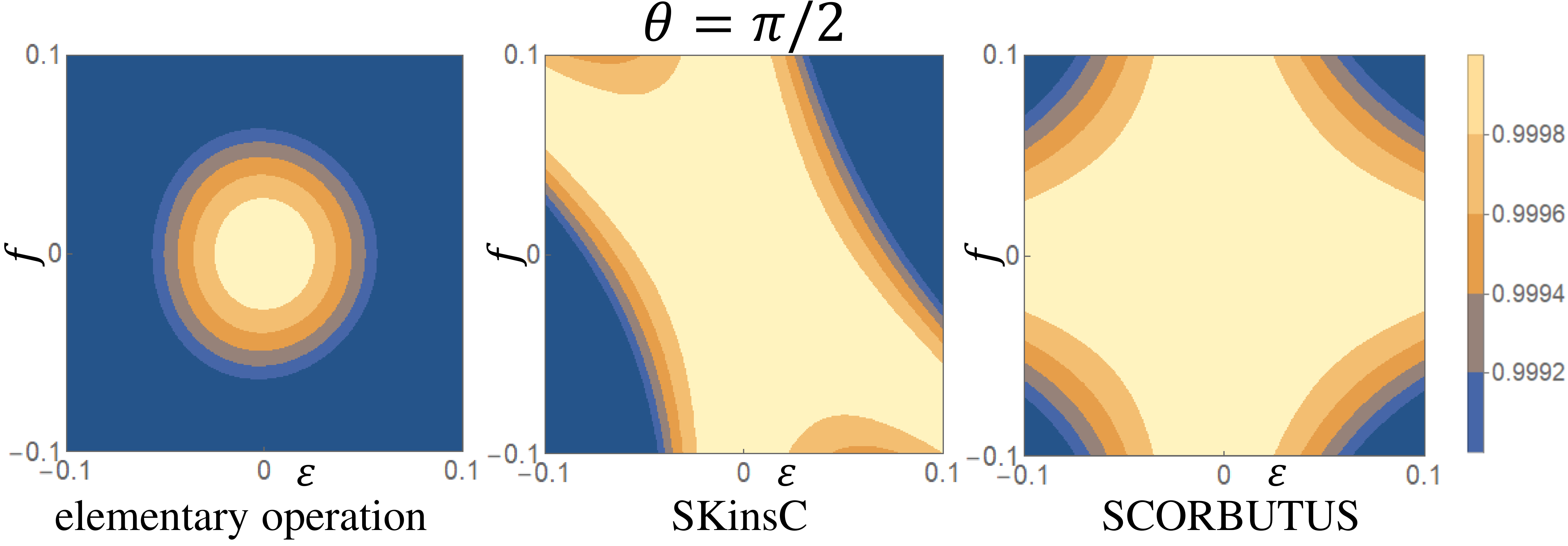}
			\caption{(color online) Gate fidelity for a target operation $(\theta)_{\phi}$ as a function of $\epsilon$ and $f$.
				We consider $\theta=\pi$ ($\theta=\pi/2$) in the upper (bottom) panels.
				\label{fig:performance}
				A brighter region indicates better gate fidelity.
			}
		\end{center}
	\end{figure*}

	We then employ the gate fidelity to verify the bi-robustness of SCORBUTUS comparing with that of the elementary operation and SKinsC.
	When a unitary operation $U$ becomes $U'$ owing to errors,
	the gate fidelity $F$ for this unitary operation under errors is defined as
	\begin{equation}
		F:=\big|{\rm tr}\bigl( U^{\dagger}U'\bigr) \big|/2.
	\end{equation}
	The gate fidelity takes a value of $0\leq F\leq 1$, and $F=1$ if and only if $U=U'$ up to a global factor.
	This quantity is often employed to evaluate the performance of CPs \cite{cummins2000use,hill2007robust,bando2012concatenated,alway2007arbitrary,JONES201191}.
	Fig.~\ref{fig:performance} depicts density plots of the gate fidelity for the elementary operation $(\theta)^{(\epsilon,f)}_{\phi}$, SKinsC, and SCORBUTUS as a function of $\epsilon$ and $f$, the magnitude of PLE and ORE, respectively.
	Evidently, SCORBUTUS exhibits stronger robustness against both errors than that of the elementary operation, as well as SKinsC.

	The proposed switchback method may also be applied to other PLE-robust CPs.
	Notably, if we adopt a higher-order PLE-robust CP, we can construct a bi-robust CP while maintaining this higher-order PLE robustness.
	Applying this method to the symmetric BB1 sequence with higher-order PLE robustness \cite{JONES201191,ichikawa2012geometric}, for example, may result in a new CP with a better accuracy, although its total operation time would be longer than that of SCORBUTUS.
	
	In summary, we constructed a CP that is robust against both PLE and ORE using the {\it switchback} technique.
	We adopted SCROFULOUS as a ``seed,'' replaced the second operation with a switchback sequence, and then added the ORE robustness by exploiting the degree of freedom resulting from this replacement.
	Thus, a $k=5$ symmetric bi-robust CP performing a rotation at an arbitrary angle $\theta$ was obtained.
	We named this sequence SCORBUTUS.
	
	Furthermore, we examined the total operation time of SCORBUTUS.
	The total operation time was found to be shorter than that of SKinsC, which has thus far been the bi-robust CP performing a rotation at an arbitrary angle $\theta$ with the shortest operation time.
	This feature of SCORBUTUS is expected to be beneficial in applications requiring short one-qubit operations.
	
	\section*{Acknowledgment}
	This work was supported by JSPS Grants-in-Aid for Scientific Research (21K03423) and CREST (JPMJCR1774).

	\begin{widetext}
		\section{Supplementary Material}
		\subsection{ORE robustness condition for symmetric CPs}
		
		Here, we explain the reason for replacing $(\pi)_{\phi_{2}}$ in SCROFULOUS with a symmetric sequence.
		We show that the ORE robustness condition for symmetric CPs provides only one equality, although it seems to exhibit a $2\times 2$ matrix equality.
		We consider a symmetric sequence $(\theta_{1})_{\phi_{1}}\cdots(\theta_{k})_{\phi_{k}}\cdots(\theta_{1})_{\phi_{1}}$.
		The first-order ORE robustness condition is given as
		\begin{align}
			& s_{1}\sigma_{z}(\theta_{2})_{\phi_{2}}\cdots(\theta_{k})_{\phi_{k}}\cdots(\theta_{1})_{\phi_{1}} \nonumber\\
			&+ s_{2} (\theta_{1})_{\phi_{1}}\sigma_{z}(\theta_{3})_{\phi_{3}} \cdots(\theta_{k})_{\phi_{k}}\cdots(\theta_{1})_{\phi_{1}}+\cdots\nonumber\\
			&+s_{k}(\theta_{1})_{\phi_{1}}\cdots(\theta_{k-1})_{\phi_{k-1}}\sigma_{z}(\theta_{k-1})_{\phi_{k-1}}\cdots(\theta_{1})_{\phi_{1}}+\cdots\nonumber\\
			&+s_{2} (\theta_{1})_{\phi_{1}}\cdots(\theta_{k})_{\phi_{k}}\cdots(\theta_{3})_{\phi_{3}}\sigma_{z} (\theta_{1})_{\phi_{1}}\nonumber\\
			&+s_{1}(\theta_{1})_{\phi_{1}}\cdots(\theta_{k})_{\phi_{k}}\cdots(\theta_{2})_{\phi_{2}}\sigma_{z})={\bold 0},
		\end{align}
		where $s_{i}:=\sin(\theta_{i}/2)$, and $\bold 0$ is the $2\times 2$ zero matrix.
		We rewrite the above equality as
		\begin{align}
			\sigma_{z}\Biggl(&s_{1}\Bigl((\theta_{2})_{\phi_{2}}\cdots(\theta_{k})_{\phi_{k}}\cdots(\theta_{1})_{\phi_{1}}+(-\theta_{1})_{\phi_{1}}\cdots(-\theta_{k})_{\phi_{k}}\cdots(-\theta_{2})_{\phi_{2}}\Bigr) +\nonumber\\
			&+ s_{2}\Bigl( (-\theta_{1})_{\phi_{1}}\sigma_{z}(\theta_{3})_{\phi_{3}} \cdots(\theta_{k})_{\phi_{k}}\cdots(\theta_{1})_{\phi_{1}}+ (-\theta_{1})_{\phi_{1}}\cdots(-\theta_{k})_{\phi_{k}}\cdots(-\theta_{3})_{\phi_{3}} (\theta_{1})_{\phi_{1}}\Bigr)+\cdots +s_{k}\sigma_{0}\Biggr)\nonumber\\
			=\sigma_{z}&\Biggl(s_{1}\Bigl((\theta_{2})_{\phi_{2}}\cdots(\theta_{k})_{\phi_{k}}\cdots(\theta_{1})_{\phi_{1}}+{\rm h.c.}\Bigr) + s_{2}\Bigl( (-\theta_{1})_{\phi_{1}}(\theta_{3})_{\phi_{3}} \cdots(\theta_{k})_{\phi_{k}}\cdots(\theta_{1})_{\phi_{1}}+{\rm h.c.}\Bigr)+\cdots+s_{k}\sigma_{0}\Biggr)
			={\bold 0},
		\end{align}
		where $\sigma_{0}$ is the $2\times 2$ identity matrix, and we use the relation $(\theta)_{\phi}\sigma_{z}=\sigma_{z}(-\theta)_{\phi}$ and $(-\theta)_{\phi}=(\theta)^{^\dagger}_{\phi}$.
		Note that $ U + U ^{\dagger}\propto \sigma_{0}$ for any $2 \times 2$ unitary matrix $U$.
		Then, we obtain the following condition:
		\begin{equation}
			(s_{1}\alpha_{1}+s_{2}\alpha_{2}+\cdots+s_{k})\sigma_{z}\sigma_{0}={\bold 0} ~\Longleftrightarrow~ s_{1}\alpha_{1}+s_{2}\alpha_{2}+\cdots+s_{k}=0,
		\end{equation}
		where $\alpha_{i}$ $(1\leq i \leq k-1)$ is defined as
		\begin{align}
			\alpha_{i}\sigma_{0}=&(-\theta_{1})_{\phi_{1}}\cdots(-\theta_{i-1})_{\phi_{i-1}}(-\theta_{i-1})_{\phi_{i-1}}\cdots (\theta_{i+1})_{\phi_{i+1}}\cdots(\theta_{k})_{\phi_{k}}\cdots  (\theta_{1})_{\phi_{1}}+ {\rm h.c.}.
		\end{align}
		Hence, we prove that the ORE robustness condition for symmetric CPs provides only one equality, although it is originally a matrix equality.
		Therefore, we replace $(\pi)_{\phi_{2}}$ in SCROFULOUS with a symmetric sequence. We only need to introduce one parameter $\theta_{r}$ to obtain the ORE robustness.
		
		\subsection{derivation of  Eq. (8)}
		
		According to the previous section, the ORE robustness condition for SCORBUTUS is given as
		\begin{equation}
			\sin(\theta_{1}/2)\alpha_{1}+\sin(\theta_{r}/2) \alpha_{r}+\sin \bigl((\pi+2 \theta_{r})/2\bigr)=0.
		\end{equation}
		The straightforward calculation shows that $\alpha_{1}$ and $\alpha_{r}$ are
		\begin{align}
			\alpha_{1}=&-2 \cos(\phi_{2}-\phi_{1}) \sin(\theta_{1}/2)=\frac{\sin (\theta_{1}/2)\pi}{\theta_{1}},\nonumber\\
			\alpha_{r}=& 2 \cos \bigl((\pi+\theta_{2})/2\bigr),
		\end{align}
		where we use the relation $\cos(\phi_{2}-\phi_{1})=-\pi/2 \theta_{1}$ of SCROFULOUS.
		The condition is finally rewritten as 
		\begin{align}
			\sin(\theta_{1}/2)\alpha_{1}&+\sin(\theta_{r}/2) \alpha_{r}+\sin \bigl((\pi+2 \theta_{r})/2\bigr)\nonumber\\
			=&~\frac{\sin^{2}(\theta_{1}/2)\pi}{\theta}+2 \sin(\theta_{r}/2)\cos \bigl((\pi+\theta_{2})/2\bigr)+\sin \bigl((\pi+2 \theta_{r})/2\bigr)\nonumber\\
			=&~\frac{\sin^{2}(\theta_{1}/2)\pi}{\theta}-2 \sin^{2}(\theta_{r}/2)+\cos (\theta_{r})\nonumber\\
			=&~\frac{\sin^{2}(\theta_{1}/2)\pi}{\theta}- 1+2 \cos(\theta_{r})=0\nonumber\\
			\Longleftrightarrow&~~\cos \theta_{r}=\frac{1}{2}\biggl(1- \frac{\pi (\sin(\theta_{1}/2))^{2}}{\theta_{1}} \biggr).
		\end{align}
		Thus, we obtain Eq. (8).
	\end{widetext}

\begin{thebibliography}{33}%
		\makeatletter
		\providecommand \@ifxundefined [1]{%
			\@ifx{#1\undefined}
		}%
		\providecommand \@ifnum [1]{%
			\ifnum #1\expandafter \@firstoftwo
			\else \expandafter \@secondoftwo
			\fi
		}%
		\providecommand \@ifx [1]{%
			\ifx #1\expandafter \@firstoftwo
			\else \expandafter \@secondoftwo
			\fi
		}%
		\providecommand \natexlab [1]{#1}%
		\providecommand \enquote  [1]{``#1''}%
		\providecommand \bibnamefont  [1]{#1}%
		\providecommand \bibfnamefont [1]{#1}%
		\providecommand \citenamefont [1]{#1}%
		\providecommand \href@noop [0]{\@secondoftwo}%
		\providecommand \href [0]{\begingroup \@sanitize@url \@href}%
		\providecommand \@href[1]{\@@startlink{#1}\@@href}%
		\providecommand \@@href[1]{\endgroup#1\@@endlink}%
		\providecommand \@sanitize@url [0]{\catcode `\\12\catcode `\$12\catcode
			`\&12\catcode `\#12\catcode `\^12\catcode `\_12\catcode `\%12\relax}%
		\providecommand \@@startlink[1]{}%
		\providecommand \@@endlink[0]{}%
		\providecommand \url  [0]{\begingroup\@sanitize@url \@url }%
		\providecommand \@url [1]{\endgroup\@href {#1}{\urlprefix }}%
		\providecommand \urlprefix  [0]{URL }%
		\providecommand \Eprint [0]{\href }%
		\providecommand \doibase [0]{https://doi.org/}%
		\providecommand \selectlanguage [0]{\@gobble}%
		\providecommand \bibinfo  [0]{\@secondoftwo}%
		\providecommand \bibfield  [0]{\@secondoftwo}%
		\providecommand \translation [1]{[#1]}%
		\providecommand \BibitemOpen [0]{}%
		\providecommand \bibitemStop [0]{}%
		\providecommand \bibitemNoStop [0]{.\EOS\space}%
		\providecommand \EOS [0]{\spacefactor3000\relax}%
		\providecommand \BibitemShut  [1]{\csname bibitem#1\endcsname}%
		\let\auto@bib@innerbib\@empty
		\bibitem [{\citenamefont {Bennett}\ and\ \citenamefont
			{DiVincenzo}(2000)}]{bennett2000quantum}%
		\BibitemOpen
		\bibfield  {author} {\bibinfo {author} {\bibfnamefont {C.~H.}\ \bibnamefont
				{Bennett}}\ and\ \bibinfo {author} {\bibfnamefont {D.~P.}\ \bibnamefont
				{DiVincenzo}},\ }\bibfield  {title} {\bibinfo {title} {Quantum information
				and computation},\ }\href@noop {} {\bibfield  {journal} {\bibinfo  {journal}
				{Nature}\ }\textbf {\bibinfo {volume} {404}},\ \bibinfo {pages} {247}
			(\bibinfo {year} {2000})}\BibitemShut {NoStop}%
		\bibitem [{\citenamefont {Nielsen}\ and\ \citenamefont
			{Chuang}(2000)}]{Nielsen2000}%
		\BibitemOpen
		\bibfield  {author} {\bibinfo {author} {\bibfnamefont {M.}~\bibnamefont
				{Nielsen}}\ and\ \bibinfo {author} {\bibfnamefont {I.}~\bibnamefont
				{Chuang}},\ }\href {https://books.google.co.jp/books?id=aai-P4V9GJ8C} {\emph
			{\bibinfo {title} {Quantum Computation and Quantum Information}}},\ Cambridge
		Series on Information and the Natural Sciences\ (\bibinfo  {publisher}
		{Cambridge University Press},\ \bibinfo {year} {2000})\BibitemShut {NoStop}%
		\bibitem [{\citenamefont {Nakahara}(2008)}]{nakahara2008quantum}%
		\BibitemOpen
		\bibfield  {author} {\bibinfo {author} {\bibfnamefont {M.}~\bibnamefont
				{Nakahara}},\ }\href@noop {} {\emph {\bibinfo {title} {Quantum computing:
					from linear algebra to physical realizations}}}\ (\bibinfo  {publisher} {CRC
			Press},\ \bibinfo {year} {2008})\BibitemShut {NoStop}%
		\bibitem [{\citenamefont {Ekert}(1991)}]{ekert1991quantum}%
		\BibitemOpen
		\bibfield  {author} {\bibinfo {author} {\bibfnamefont {A.~K.}\ \bibnamefont
				{Ekert}},\ }\bibfield  {title} {\bibinfo {title} {Quantum cryptography based
				on Bell’s theorem},\ }\href@noop {} {\bibfield  {journal} {\bibinfo
				{journal} {Physical Review Letters}\ }\textbf {\bibinfo {volume} {67}},\
			\bibinfo {pages} {661} (\bibinfo {year} {1991})}\BibitemShut {NoStop}%
		\bibitem [{\citenamefont {Gisin}\ and\ \citenamefont
			{Thew}(2007)}]{gisin2007quantum}%
		\BibitemOpen
		\bibfield  {author} {\bibinfo {author} {\bibfnamefont {N.}~\bibnamefont
				{Gisin}}\ and\ \bibinfo {author} {\bibfnamefont {R.}~\bibnamefont {Thew}},\
		}\bibfield  {title} {\bibinfo {title} {Quantum communication},\ }\href@noop
		{} {\bibfield  {journal} {\bibinfo  {journal} {Nature Photonics}\ }\textbf
			{\bibinfo {volume} {1}},\ \bibinfo {pages} {165} (\bibinfo {year}
			{2007})}\BibitemShut {NoStop}%
		\bibitem [{\citenamefont {Chen}\ \emph {et~al.}(2021)\citenamefont {Chen},
			\citenamefont {Zhang}, \citenamefont {Chen}, \citenamefont {Cai},
			\citenamefont {Liao}, \citenamefont {Zhang}, \citenamefont {Chen},
			\citenamefont {Yin}, \citenamefont {Ren}, \citenamefont {Chen} \emph
			{et~al.}}]{chen2021integrated}%
		\BibitemOpen
		\bibfield  {author} {\bibinfo {author} {\bibfnamefont {Y.-A.}\ \bibnamefont
				{Chen}}, \bibinfo {author} {\bibfnamefont {Q.}~\bibnamefont {Zhang}},
			\bibinfo {author} {\bibfnamefont {T.-Y.}\ \bibnamefont {Chen}}, \bibinfo
			{author} {\bibfnamefont {W.-Q.}\ \bibnamefont {Cai}}, \bibinfo {author}
			{\bibfnamefont {S.-K.}\ \bibnamefont {Liao}}, \bibinfo {author}
			{\bibfnamefont {J.}~\bibnamefont {Zhang}}, \bibinfo {author} {\bibfnamefont
				{K.}~\bibnamefont {Chen}}, \bibinfo {author} {\bibfnamefont {J.}~\bibnamefont
				{Yin}}, \bibinfo {author} {\bibfnamefont {J.-G.}\ \bibnamefont {Ren}},
			\bibinfo {author} {\bibfnamefont {Z.}~\bibnamefont {Chen}}, \emph {et~al.},\
		}\bibfield  {title} {\bibinfo {title} {An integrated space-to-ground quantum
				communication network over 4,600 kilometres},\ }\href@noop {} {\bibfield
			{journal} {\bibinfo  {journal} {Nature}\ }\textbf {\bibinfo {volume} {589}},\
			\bibinfo {pages} {214} (\bibinfo {year} {2021})}\BibitemShut {NoStop}%
		\bibitem [{\citenamefont {Helstrom}(1976)}]{helstrom1976quantum}%
		\BibitemOpen
		\bibfield  {author} {\bibinfo {author} {\bibfnamefont {C.~W.}\ \bibnamefont
				{Helstrom}},\ }\href@noop {} {\emph {\bibinfo {title} {Quantum detection and
					estimation theory}}},\ Vol.~\bibinfo {volume} {84}\ (\bibinfo  {publisher}
		{Academic Press New York},\ \bibinfo {year} {1976})\BibitemShut {NoStop}%
		\bibitem [{\citenamefont {Caves}(1981)}]{caves1981quantum}%
		\BibitemOpen
		\bibfield  {author} {\bibinfo {author} {\bibfnamefont {C.~M.}\ \bibnamefont
				{Caves}},\ }\bibfield  {title} {\bibinfo {title} {Quantum-mechanical noise in
				an interferometer},\ }\href@noop {} {\bibfield  {journal} {\bibinfo
				{journal} {Physical Review D}\ }\textbf {\bibinfo {volume} {23}},\ \bibinfo
			{pages} {1693} (\bibinfo {year} {1981})}\BibitemShut {NoStop}%
		\bibitem [{\citenamefont {Holevo}(2011)}]{holevo2011probabilistic}%
		\BibitemOpen
		\bibfield  {author} {\bibinfo {author} {\bibfnamefont {A.~S.}\ \bibnamefont
				{Holevo}},\ }\href@noop {} {\emph {\bibinfo {title} {Probabilistic and
					statistical aspects of quantum theory}}},\ Vol.~\bibinfo {volume} {1}\
		(\bibinfo  {publisher} {Springer Science \& Business Media},\ \bibinfo {year}
		{2011})\BibitemShut {NoStop}%
		\bibitem [{\citenamefont {Counsell}\ \emph {et~al.}(1985)\citenamefont
			{Counsell}, \citenamefont {Levitt},\ and\ \citenamefont
			{Ernst}}]{counsell1985analytical}%
		\BibitemOpen
		\bibfield  {author} {\bibinfo {author} {\bibfnamefont {C.}~\bibnamefont
				{Counsell}}, \bibinfo {author} {\bibfnamefont {M.}~\bibnamefont {Levitt}},\
			and\ \bibinfo {author} {\bibfnamefont {R.}~\bibnamefont {Ernst}},\ }\bibfield
		{title} {\bibinfo {title} {Analytical theory of composite pulses},\
		}\href@noop {} {\bibfield  {journal} {\bibinfo  {journal} {Journal of
					Magnetic Resonance}\ }\textbf {\bibinfo {volume} {63}},\ \bibinfo
			{pages} {133} (\bibinfo {year} {1985})}\BibitemShut {NoStop}%
		\bibitem [{\citenamefont {Levitt}(1986)}]{levitt1986composite}%
		\BibitemOpen
		\bibfield  {author} {\bibinfo {author} {\bibfnamefont {M.~H.}\ \bibnamefont
				{Levitt}},\ }\bibfield  {title} {\bibinfo {title} {Composite pulses},\
		}\href@noop {} {\bibfield  {journal} {\bibinfo  {journal} {Progress in
					Nuclear Magnetic Resonance Spectroscopy}\ }\textbf {\bibinfo {volume} {18}},\
			\bibinfo {pages} {61} (\bibinfo {year} {1986})}\BibitemShut {NoStop}%
		\bibitem [{\citenamefont {Claridge}(2016)}]{claridge2016high}%
		\BibitemOpen
		\bibfield  {author} {\bibinfo {author} {\bibfnamefont {T.~D.}\ \bibnamefont
				{Claridge}},\ }\href@noop {} {\emph {\bibinfo {title} {High-resolution NMR
					techniques in organic chemistry}}},\ Vol.~\bibinfo {volume} {27}\ (\bibinfo
		{publisher} {Elsevier},\ \bibinfo {year} {2016})\BibitemShut {NoStop}%
		\bibitem [{\citenamefont {Levitt}(2013)}]{levitt2013spin}%
		\BibitemOpen
		\bibfield  {author} {\bibinfo {author} {\bibfnamefont {M.~H.}\ \bibnamefont
				{Levitt}},\ }\href@noop {} {\emph {\bibinfo {title} {Spin dynamics: basics of
					nuclear magnetic resonance}}}\ (\bibinfo  {publisher} {John Wiley \& Sons},\
		\bibinfo {year} {2013})\BibitemShut {NoStop}%
		\bibitem [{\citenamefont {Gershenfeld}\ and\ \citenamefont
			{Chuang}(1997)}]{gershenfeld1997bulk}%
		\BibitemOpen
		\bibfield  {author} {\bibinfo {author} {\bibfnamefont {N.~A.}\ \bibnamefont
				{Gershenfeld}}\ and\ \bibinfo {author} {\bibfnamefont {I.~L.}\ \bibnamefont
				{Chuang}},\ }\bibfield  {title} {\bibinfo {title} {Bulk spin-resonance
				quantum computation},\ }\href@noop {} {\bibfield  {journal} {\bibinfo
				{journal} {Science}\ }\textbf {\bibinfo {volume} {275}},\ \bibinfo {pages}
			{350} (\bibinfo {year} {1997})}\BibitemShut {NoStop}%
		\bibitem [{\citenamefont {Cory}\ \emph {et~al.}(1997)\citenamefont {Cory},
			\citenamefont {Fahmy},\ and\ \citenamefont {Havel}}]{cory1997ensemble}%
		\BibitemOpen
		\bibfield  {author} {\bibinfo {author} {\bibfnamefont {D.~G.}\ \bibnamefont
				{Cory}}, \bibinfo {author} {\bibfnamefont {A.~F.}\ \bibnamefont {Fahmy}},\
			and\ \bibinfo {author} {\bibfnamefont {T.~F.}\ \bibnamefont {Havel}},\
		}\bibfield  {title} {\bibinfo {title} {Ensemble quantum computing by NMR
				spectroscopy},\ }\href@noop {} {\bibfield  {journal} {\bibinfo  {journal}
				{Proceedings of the National Academy of Sciences}\ }\textbf {\bibinfo
				{volume} {94}},\ \bibinfo {pages} {1634} (\bibinfo {year}
			{1997})}\BibitemShut {NoStop}%
		\bibitem [{\citenamefont {Vandersypen}\ \emph {et~al.}(2001)\citenamefont
			{Vandersypen}, \citenamefont {Steffen}, \citenamefont {Breyta}, \citenamefont
			{Yannoni}, \citenamefont {Sherwood},\ and\ \citenamefont
			{Chuang}}]{vandersypen2001experimental}%
		\BibitemOpen
		\bibfield  {author} {\bibinfo {author} {\bibfnamefont {L.~M.}\ \bibnamefont
				{Vandersypen}}, \bibinfo {author} {\bibfnamefont {M.}~\bibnamefont
				{Steffen}}, \bibinfo {author} {\bibfnamefont {G.}~\bibnamefont {Breyta}},
			\bibinfo {author} {\bibfnamefont {C.~S.}\ \bibnamefont {Yannoni}}, \bibinfo
			{author} {\bibfnamefont {M.~H.}\ \bibnamefont {Sherwood}},\ and\ \bibinfo
			{author} {\bibfnamefont {I.~L.}\ \bibnamefont {Chuang}},\ }\bibfield  {title}
		{\bibinfo {title} {Experimental realization of Shor's quantum factoring
				algorithm using nuclear magnetic resonance},\ }\href@noop {} {\bibfield
			{journal} {\bibinfo  {journal} {Nature}\ }\textbf {\bibinfo {volume} {414}},\
			\bibinfo {pages} {883} (\bibinfo {year} {2001})}\BibitemShut {NoStop}%
		\bibitem [{\citenamefont {Jones}(2011)}]{JONES201191}%
		\BibitemOpen
		\bibfield  {author} {\bibinfo {author} {\bibfnamefont {J.~A.}\ \bibnamefont
				{Jones}},\ }\bibfield  {title} {\bibinfo {title} {Quantum computing with
				nmr},\ }\href {https://doi.org/https://doi.org/10.1016/j.pnmrs.2010.11.001}
		{\bibfield  {journal} {\bibinfo  {journal} {Progress in Nuclear Magnetic
					Resonance Spectroscopy}\ }\textbf {\bibinfo {volume} {59}},\ \bibinfo {pages}
			{91} (\bibinfo {year} {2011})}\BibitemShut {NoStop}%
		\bibitem [{\citenamefont {Bando}\ \emph {et~al.}(2020)\citenamefont {Bando},
			\citenamefont {Ichikawa}, \citenamefont {Kondo}, \citenamefont {Nemoto},
			\citenamefont {Nakahara},\ and\ \citenamefont
			{Shikano}}]{bando2020concatenated}%
		\BibitemOpen
		\bibfield  {author} {\bibinfo {author} {\bibfnamefont {M.}~\bibnamefont
				{Bando}}, \bibinfo {author} {\bibfnamefont {T.}~\bibnamefont {Ichikawa}},
			\bibinfo {author} {\bibfnamefont {Y.}~\bibnamefont {Kondo}}, \bibinfo
			{author} {\bibfnamefont {N.}~\bibnamefont {Nemoto}}, \bibinfo {author}
			{\bibfnamefont {M.}~\bibnamefont {Nakahara}},\ and\ \bibinfo {author}
			{\bibfnamefont {Y.}~\bibnamefont {Shikano}},\ }\bibfield  {title} {\bibinfo
			{title} {Concatenated composite pulses applied to liquid-state nuclear
				magnetic resonance spectroscopy},\ }\href@noop {} {\bibfield  {journal}
			{\bibinfo  {journal} {Scientific Reports}\ }\textbf {\bibinfo {volume}
				{10}},\ \bibinfo {pages} {1} (\bibinfo {year} {2020})}\BibitemShut {NoStop}%
		\bibitem [{\citenamefont {Said}\ and\ \citenamefont
			{Twamley}(2009)}]{said2009robust}%
		\BibitemOpen
		\bibfield  {author} {\bibinfo {author} {\bibfnamefont {R.}~\bibnamefont
				{Said}}\ and\ \bibinfo {author} {\bibfnamefont {J.}~\bibnamefont {Twamley}},\
		}\bibfield  {title} {\bibinfo {title} {Robust control of entanglement in a
				nitrogen-vacancy center coupled to a c 13 nuclear spin in diamond},\
		}\href@noop {} {\bibfield  {journal} {\bibinfo  {journal} {Physical Review
					A}\ }\textbf {\bibinfo {volume} {80}},\ \bibinfo {pages} {032303} (\bibinfo
			{year} {2009})}\BibitemShut {NoStop}%
		\bibitem [{\citenamefont {Collin}\ \emph {et~al.}(2004)\citenamefont {Collin},
			\citenamefont {Ithier}, \citenamefont {Aassime}, \citenamefont {Joyez},
			\citenamefont {Vion},\ and\ \citenamefont {Esteve}}]{collin2004nmr}%
		\BibitemOpen
		\bibfield  {author} {\bibinfo {author} {\bibfnamefont {E.}~\bibnamefont
				{Collin}}, \bibinfo {author} {\bibfnamefont {G.}~\bibnamefont {Ithier}},
			\bibinfo {author} {\bibfnamefont {A.}~\bibnamefont {Aassime}}, \bibinfo
			{author} {\bibfnamefont {P.}~\bibnamefont {Joyez}}, \bibinfo {author}
			{\bibfnamefont {D.}~\bibnamefont {Vion}},\ and\ \bibinfo {author}
			{\bibfnamefont {D.}~\bibnamefont {Esteve}},\ }\bibfield  {title} {\bibinfo
			{title} {Nmr-like control of a quantum bit superconducting circuit},\
		}\href@noop {} {\bibfield  {journal} {\bibinfo  {journal} {Physical Review
					Letters}\ }\textbf {\bibinfo {volume} {93}},\ \bibinfo {pages} {157005}
			(\bibinfo {year} {2004})}\BibitemShut {NoStop}%
		\bibitem [{\citenamefont {Gulde}\ \emph {et~al.}(2003)\citenamefont {Gulde},
			\citenamefont {Riebe}, \citenamefont {Lancaster}, \citenamefont {Becher},
			\citenamefont {Eschner}, \citenamefont {H{\"a}ffner}, \citenamefont
			{Schmidt-Kaler}, \citenamefont {Chuang},\ and\ \citenamefont
			{Blatt}}]{gulde2003implementation}%
		\BibitemOpen
		\bibfield  {author} {\bibinfo {author} {\bibfnamefont {S.}~\bibnamefont
				{Gulde}}, \bibinfo {author} {\bibfnamefont {M.}~\bibnamefont {Riebe}},
			\bibinfo {author} {\bibfnamefont {G.~P.}\ \bibnamefont {Lancaster}}, \bibinfo
			{author} {\bibfnamefont {C.}~\bibnamefont {Becher}}, \bibinfo {author}
			{\bibfnamefont {J.}~\bibnamefont {Eschner}}, \bibinfo {author} {\bibfnamefont
				{H.}~\bibnamefont {H{\"a}ffner}}, \bibinfo {author} {\bibfnamefont
				{F.}~\bibnamefont {Schmidt-Kaler}}, \bibinfo {author} {\bibfnamefont {I.~L.}\
				\bibnamefont {Chuang}},\ and\ \bibinfo {author} {\bibfnamefont
				{R.}~\bibnamefont {Blatt}},\ }\bibfield  {title} {\bibinfo {title}
			{Implementation of the deutsch--jozsa algorithm on an ion-trap quantum
				computer},\ }\href@noop {} {\bibfield  {journal} {\bibinfo  {journal}
				{Nature}\ }\textbf {\bibinfo {volume} {421}},\ \bibinfo {pages} {48}
			(\bibinfo {year} {2003})}\BibitemShut {NoStop}%
		\bibitem [{\citenamefont {Timoney}\ \emph {et~al.}(2008)\citenamefont
			{Timoney}, \citenamefont {Elman}, \citenamefont {Glaser}, \citenamefont
			{Weiss}, \citenamefont {Johanning}, \citenamefont {Neuhauser},\ and\
			\citenamefont {Wunderlich}}]{timoney2008error}%
		\BibitemOpen
		\bibfield  {author} {\bibinfo {author} {\bibfnamefont {N.}~\bibnamefont
				{Timoney}}, \bibinfo {author} {\bibfnamefont {V.}~\bibnamefont {Elman}},
			\bibinfo {author} {\bibfnamefont {S.}~\bibnamefont {Glaser}}, \bibinfo
			{author} {\bibfnamefont {C.}~\bibnamefont {Weiss}}, \bibinfo {author}
			{\bibfnamefont {M.}~\bibnamefont {Johanning}}, \bibinfo {author}
			{\bibfnamefont {W.}~\bibnamefont {Neuhauser}},\ and\ \bibinfo {author}
			{\bibfnamefont {C.}~\bibnamefont {Wunderlich}},\ }\bibfield  {title}
		{\bibinfo {title} {Error-resistant single-qubit gates with trapped ions},\
		}\href@noop {} {\bibfield  {journal} {\bibinfo  {journal} {Physical Review
					A}\ }\textbf {\bibinfo {volume} {77}},\ \bibinfo {pages} {052334} (\bibinfo
			{year} {2008})}\BibitemShut {NoStop}%
		\bibitem [{\citenamefont {Brown}\ \emph {et~al.}(2004)\citenamefont {Brown},
			\citenamefont {Harrow},\ and\ \citenamefont {Chuang}}]{brown2004arbitrarily}%
		\BibitemOpen
		\bibfield  {author} {\bibinfo {author} {\bibfnamefont {K.~R.}\ \bibnamefont
				{Brown}}, \bibinfo {author} {\bibfnamefont {A.~W.}\ \bibnamefont {Harrow}},\
			and\ \bibinfo {author} {\bibfnamefont {I.~L.}\ \bibnamefont {Chuang}},\
		}\bibfield  {title} {\bibinfo {title} {Arbitrarily accurate composite pulse
				sequences},\ }\href@noop {} {\bibfield  {journal} {\bibinfo  {journal}
				{Physical Review A}\ }\textbf {\bibinfo {volume} {70}},\ \bibinfo {pages}
			{052318} (\bibinfo {year} {2004})}\BibitemShut {NoStop}%
		\bibitem [{\citenamefont {Wimperis}(1994)}]{wimperis1994broadband}%
		\BibitemOpen
		\bibfield  {author} {\bibinfo {author} {\bibfnamefont {S.}~\bibnamefont
				{Wimperis}},\ }\bibfield  {title} {\bibinfo {title} {Broadband, narrowband,
				and passband composite pulses for use in advanced NMR experiments},\
		}\href@noop {} {\bibfield  {journal} {\bibinfo  {journal} {Journal of
					Magnetic Resonance, Series A}\ }\textbf {\bibinfo {volume} {109}},\ \bibinfo
			{pages} {221} (\bibinfo {year} {1994})}\BibitemShut {NoStop}%
		\bibitem [{\citenamefont {Cummins}\ \emph {et~al.}(2003)\citenamefont
			{Cummins}, \citenamefont {Llewellyn},\ and\ \citenamefont
			{Jones}}]{cummins2003tackling}%
		\BibitemOpen
		\bibfield  {author} {\bibinfo {author} {\bibfnamefont {H.~K.}\ \bibnamefont
				{Cummins}}, \bibinfo {author} {\bibfnamefont {G.}~\bibnamefont {Llewellyn}},\
			and\ \bibinfo {author} {\bibfnamefont {J.~A.}\ \bibnamefont {Jones}},\
		}\bibfield  {title} {\bibinfo {title} {Tackling systematic errors in quantum
				logic gates with composite rotations},\ }\href@noop {} {\bibfield  {journal}
			{\bibinfo  {journal} {Physical Review A}\ }\textbf {\bibinfo {volume} {67}},\
			\bibinfo {pages} {042308} (\bibinfo {year} {2003})}\BibitemShut {NoStop}%
		\bibitem [{\citenamefont {Cummins}\ and\ \citenamefont
			{Jones}(2000)}]{cummins2000use}%
		\BibitemOpen
		\bibfield  {author} {\bibinfo {author} {\bibfnamefont {H.}~\bibnamefont
				{Cummins}}\ and\ \bibinfo {author} {\bibfnamefont {J.}~\bibnamefont
				{Jones}},\ }\bibfield  {title} {\bibinfo {title} {Use of composite rotations
				to correct systematic errors in nmr quantum computation},\ }\href@noop {}
		{\bibfield  {journal} {\bibinfo  {journal} {New Journal of Physics}\ }\textbf
			{\bibinfo {volume} {2}},\ \bibinfo {pages} {6} (\bibinfo {year}
			{2000})}\BibitemShut {NoStop}%
		\bibitem [{\citenamefont {M{\"o}tt{\"o}nen}\ \emph {et~al.}(2006)\citenamefont
			{M{\"o}tt{\"o}nen}, \citenamefont {de~Sousa}, \citenamefont {Zhang},\ and\
			\citenamefont {Whaley}}]{mottonen2006high}%
		\BibitemOpen
		\bibfield  {author} {\bibinfo {author} {\bibfnamefont {M.}~\bibnamefont
				{M{\"o}tt{\"o}nen}}, \bibinfo {author} {\bibfnamefont {R.}~\bibnamefont
				{de~Sousa}}, \bibinfo {author} {\bibfnamefont {J.}~\bibnamefont {Zhang}},\
			and\ \bibinfo {author} {\bibfnamefont {K.~B.}\ \bibnamefont {Whaley}},\
		}\bibfield  {title} {\bibinfo {title} {High-fidelity one-qubit operations
				under random telegraph noise},\ }\href@noop {} {\bibfield  {journal}
			{\bibinfo  {journal} {Physical Review A}\ }\textbf {\bibinfo {volume} {73}},\
			\bibinfo {pages} {022332} (\bibinfo {year} {2006})}\BibitemShut {NoStop}%
		\bibitem [{\citenamefont {Bando}\ \emph {et~al.}(2012)\citenamefont {Bando},
			\citenamefont {Ichikawa}, \citenamefont {Kondo},\ and\ \citenamefont
			{Nakahara}}]{bando2012concatenated}%
		\BibitemOpen
		\bibfield  {author} {\bibinfo {author} {\bibfnamefont {M.}~\bibnamefont
				{Bando}}, \bibinfo {author} {\bibfnamefont {T.}~\bibnamefont {Ichikawa}},
			\bibinfo {author} {\bibfnamefont {Y.}~\bibnamefont {Kondo}},\ and\ \bibinfo
			{author} {\bibfnamefont {M.}~\bibnamefont {Nakahara}},\ }\bibfield  {title}
		{\bibinfo {title} {Concatenated composite pulses compensating simultaneous
				systematic errors},\ }\href@noop {} {\bibfield  {journal} {\bibinfo
				{journal} {Journal of the Physical Society of Japan}\ }\textbf {\bibinfo
				{volume} {82}},\ \bibinfo {pages} {014004} (\bibinfo {year}
			{2012})}\BibitemShut {NoStop}%
		\bibitem [{\citenamefont {Jones}(2013)}]{jones2013designing}%
		\BibitemOpen
		\bibfield  {author} {\bibinfo {author} {\bibfnamefont {J.~A.}\ \bibnamefont
				{Jones}},\ }\bibfield  {title} {\bibinfo {title} {Designing short robust not
				gates for quantum computation},\ }\href@noop {} {\bibfield  {journal}
			{\bibinfo  {journal} {Physical Review A}\ }\textbf {\bibinfo {volume} {87}},\
			\bibinfo {pages} {052317} (\bibinfo {year} {2013})}\BibitemShut {NoStop}%
		\bibitem [{\citenamefont {Ryan}\ \emph {et~al.}(2010)\citenamefont {Ryan},
			\citenamefont {Hodges},\ and\ \citenamefont {Cory}}]{ryan2010robust}%
		\BibitemOpen
		\bibfield  {author} {\bibinfo {author} {\bibfnamefont {C.~A.}\ \bibnamefont
				{Ryan}}, \bibinfo {author} {\bibfnamefont {J.~S.}\ \bibnamefont {Hodges}},\
			and\ \bibinfo {author} {\bibfnamefont {D.~G.}\ \bibnamefont {Cory}},\
		}\bibfield  {title} {\bibinfo {title} {Robust decoupling techniques to extend
				quantum coherence in diamond},\ }\href@noop {} {\bibfield  {journal}
			{\bibinfo  {journal} {Physical Review Letters}\ }\textbf {\bibinfo {volume}
				{105}},\ \bibinfo {pages} {200402} (\bibinfo {year} {2010})}\BibitemShut
		{NoStop}%
		\bibitem [{\citenamefont {Hill}(2007)}]{hill2007robust}%
		\BibitemOpen
		\bibfield  {author} {\bibinfo {author} {\bibfnamefont {C.~D.}\ \bibnamefont
				{Hill}},\ }\bibfield  {title} {\bibinfo {title} {Robust controlled-not gates
				from almost any interaction},\ }\href@noop {} {\bibfield  {journal} {\bibinfo
				{journal} {Physical Review Letters}\ }\textbf {\bibinfo {volume} {98}},\
			\bibinfo {pages} {180501} (\bibinfo {year} {2007})}\BibitemShut {NoStop}%
		\bibitem [{\citenamefont {Alway}\ and\ \citenamefont
			{Jones}(2007)}]{alway2007arbitrary}%
		\BibitemOpen
		\bibfield  {author} {\bibinfo {author} {\bibfnamefont {W.~G.}\ \bibnamefont
				{Alway}}\ and\ \bibinfo {author} {\bibfnamefont {J.~A.}\ \bibnamefont
				{Jones}},\ }\bibfield  {title} {\bibinfo {title} {Arbitrary precision
				composite pulses for nmr quantum computing},\ }\href@noop {} {\bibfield
			{journal} {\bibinfo  {journal} {Journal of Magnetic Resonance}\ }\textbf
			{\bibinfo {volume} {189}},\ \bibinfo {pages} {114} (\bibinfo {year}
			{2007})}\BibitemShut {NoStop}%
		\bibitem [{\citenamefont {Ichikawa}\ \emph {et~al.}(2012)\citenamefont
			{Ichikawa}, \citenamefont {Bando}, \citenamefont {Kondo},\ and\ \citenamefont
			{Nakahara}}]{ichikawa2012geometric}%
		\BibitemOpen
		\bibfield  {author} {\bibinfo {author} {\bibfnamefont {T.}~\bibnamefont
				{Ichikawa}}, \bibinfo {author} {\bibfnamefont {M.}~\bibnamefont {Bando}},
			\bibinfo {author} {\bibfnamefont {Y.}~\bibnamefont {Kondo}},\ and\ \bibinfo
			{author} {\bibfnamefont {M.}~\bibnamefont {Nakahara}},\ }\bibfield  {title}
		{\bibinfo {title} {Geometric aspects of composite pulses},\ }\href@noop {}
		{\bibfield  {journal} {\bibinfo  {journal} {Philosophical Transactions of the
					Royal Society A: Mathematical, Physical and Engineering Sciences}\ }\textbf
			{\bibinfo {volume} {370}},\ \bibinfo {pages} {4671} (\bibinfo {year}
			{2012})}\BibitemShut {NoStop}%
	\end{thebibliography}
	\end{document}